\documentclass[longauth]{aa}  
\usepackage{graphicx}
\usepackage{txfonts}
\usepackage{hyperref}
\hypersetup{
    colorlinks=true,
    linkcolor=blue,
    citecolor=blue,
    filecolor=magenta,      
    urlcolor=cyan,
    pdftitle={GN-z11 Lya},
    pdfpagemode=FullScreen,
    }

\begin{document}

   \title{GN-z11: The environment of an AGN at $z=$10.603
    }
%other possible titles : 
% The environment of 
   \subtitle{ New insights on the most distant Ly$\alpha$ detection}
   
   %%=============================================================%%
%% Prefix	-> \pfx{Dr}
%% GivenName	-> \fnm{Joergen W.}
%% Particle	-> \spfx{van der} -> surname prefix
%% FamilyName	-> \sur{Ploeg}
%% Suffix	-> \sfx{IV}
%% NatureName	-> \tanm{Poet Laureate} -> Title after name
%% Degrees	-> \dgr{MSc, PhD}
%% \author*[1,2]{\pfx{Dr} \fnm{Joergen W.} \spfx{van der} \sur{Ploeg} \sfx{IV} \tanm{Poet Laureate} 
%%                 \dgr{MSc, PhD}}\email{iauthor@gmail.com}
%%=============================================================%%
%Tier1
%Tier1

%%=============================================================%%
%% Prefix	-> \pfx{Dr}
%% GivenName	-> \fnm{Joergen W.}
%% Particle	-> \spfx{van der} -> surname prefix
%% FamilyName	-> \sur{Ploeg}
%% Suffix	-> \sfx{IV}
%% NatureName	-> \tanm{Poet Laureate} -> Title after name
%% Degrees	-> \dgr{MSc, PhD}
%% \author*[1,2]{\pfx{Dr} \fnm{Joergen W.} \spfx{van der} \sur{Ploeg} \sfx{IV} \tanm{Poet Laureate} 
%%                 \dgr{MSc, PhD}}\email{iauthor@gmail.com}
%%=============================================================%%
%Tier1
\author{
Jan Scholtz
\inst{1,2}\fnmsep\thanks{Corresponding authors: js2685@cam.ac.uk, cw795@cam.ac.uk \\ These authors contributed equally to this work.}
\and
Callum Witten
\inst{3,1}\fnmsep$^\star$
\and
Nicolas Laporte
\inst{1,2,15}
\and
Hannah \"Ubler
\inst{1,2}
\and
Michele Perna
\inst{4}
\and
Roberto Maiolino
\inst{1,2,5}
\and
Santiago Arribas
\inst{4}
\and
William M. Baker
\inst{1,2}
\and
Jake S. Bennett
\inst{7}
\and
Francesco D'Eugenio
\inst{1,2}
\and
Charlotte Simmonds
\inst{1,2} 
\and
Sandro Tacchella
\inst{1,2}
\and
Joris Witstok
\inst{1,2}
\and
Andrew J. Bunker 
\inst{8} 
\and
Stefano Carniani
\inst{6}
%Tier2 (rest alphabetical)
\and
Stéphane Charlot
\inst{9} 
\and
Giovanni Cresci
\and
Emma Curtis-Lake
\inst{10} 
\and
Daniel J.\ Eisenstein
\inst{7}
\and
Nimisha Kumari
\inst{16}
\and
Brant Robertson
\inst{11} 
\and
Bruno Rodríguez Del Pino
\inst{4} 
\and
Renske Smit
\inst{12} 
\and
Giacomo Venturi
\inst{6} 
\and
Christina C. Williams
\inst{13} 
\and
Christopher N. A. Willmer
\inst{14} 
}

\institute{
Kavli Institute for Cosmology, University of Cambridge, Madingley Road, Cambridge, 
CB3 0HA, UK\\
\and
Cavendish Laboratory, University of Cambridge, 19 JJ Thomson Avenue, Cambridge CB3 0HE, UK\\
\and
Institute of Astronomy, University of Cambridge, Madingley Road, Cambridge CB3 0HA, UK\\
\and
Centro de Astrobiolog\'ia (CAB), CSIC–INTA, Cra. de Ajalvir Km.~4, 28850- Torrej\'on de Ardoz, Madrid, Spain\\
\and
Department of Physics and Astronomy, University College London, Gower Street, London WC1E 6BT, UK\\
\and
Scuola Normale Superiore, Piazza dei Cavalieri 7, I-56126 Pisa, Italy\\
\and
Harvard University, Center for Astrophysics $|$ Harvard \& Smithsonian, 60 Garden St., Cambridge 2138, USA\\
\and
Department of Physics, University of Oxford, Denys Wilkinson Building, Keble Road, Oxford OX1 3RH, UK\\
\and
Sorbonne Universit\'e, CNRS, UMR 7095, Institut d'Astrophysique de Paris, 98 bis bd Arago, 75014 Paris, France\\
\and
Centre for Astrophysics Research, Department of Physics, Astronomy and Mathematics, University of Hertfordshire, Hatfield AL10 9AB, UK\\
\and
Department of Astronomy and Astrophysics, University of California, Santa Cruz, 1156 High Street, Santa Cruz, CA 95064, USA\\
\and
Astrophysics Research Institute, Liverpool John Moores University, 146 Brownlow Hill, Liverpool L3 5RF, UK\\
\and
NSF’s National Optical-Infrared Astronomy Research Laboratory, 950 North Cherry Avenue, Tucson, AZ 85719, USA\\
\and
Steward Observatory University of Arizona 933 N. Cherry Avenue Tucson AZ 85721, USA\\
\and
Aix Marseille Université, CNRS, CNES, LAM (Laboratoire d’Astrophysique de Marseille), UMR 7326, 13388 Marseille, France \\
\and
AURA for European Space Agency, Space Telescope Science Institute, 3700 San Martin Drive. Baltimore, MD, 21210\\
}

   \authorrunning{Scholtz and Witten et al.}
   \date{}

%%==================================%%
%% sample for unstructured abstract %%
%%==================================%%
 
  \abstract
  % context heading (optional)
  % {} leave it empty if necessary  
   {Recent observations with the \textit{James Webb} Space Telescope (JWST) have further refined the spectroscopic redshift of GN-z11, one of the most distant galaxies identified with the \textit{Hubble} Space Telescope (HST) at $z=10.603$. The presence of extremely dense gas ($>10^{10}$ cm$^{-3}$), the detection of high-ionisation lines and of CII*1335 emission, as well as the presence of an ionisation cone, indicate that GN-z11 also hosts an Active Galactic Nucleus (AGN). Further photometric and spectroscopic follow-up demonstrates that it lies in a large-scale, overdense structure with possible signatures of Population III (PopIII) stars in its halo. Surprisingly, Ly$\alpha$ has also been detected despite the expected largely neutral inter-galactic medium at such a redshift. We exploit recent JWST/NIRSpec IFU observations to demonstrate that the Ly$\alpha$ emission in GN-z11 is part of an extended halo with a minimum size of 0.8--3.2 kpc, depending on the definition used to derive the halo size. The surface brightness of the Ly$\alpha$ halo around GN-z11 appears consistent with Ly$\alpha$ halos observed around $z\sim6$ quasars. At the wavelength of Ly$\alpha$ at $z\sim$10.6, we identify three other emission line candidates within the IFU Field-of-View with no UV rest-frame counterpart visible in deep images from the JWST/NIRCam. If confirmed, this could be the first evidence that the local region of GN-z11 represents a candidate protocluster core, forming just 400 Myr after the Big Bang. We give a first estimate of the dark matter halo mass of this structure ($M_h$=2.96$^{+0.44}_{-0.39} \times$10$^{10}$ M$_{\odot}$), consistent with a Coma-like cluster progenitor.
}


   \maketitle
%
%-------------------------------------------------------------------

\section{Introduction}

One of the most intriguing challenges of modern extragalactic astronomy is to understand how the first luminous objects emerged from a dark and neutral Universe less than 300 million years after the Big Bang. This quest for \textit{Cosmic Dawn} has been a scientific objective of many telescopes and instruments since the 1950s (for a review see \citealt{2022wgwb.book.....E}). Deep and large photometric near infra-red surveys have been undertaken to identify galaxies during the Epoch of Reionisation (EoR), at $z\geq$6, using the Lyman Break technique \citep{1996ApJ...462L..17S}, a well-known and tested method based on the fact that UV photons emitted at energies higher than Ly$\alpha$ ($\lambda \leq$1216\AA) are almost completely absorbed by the neutral gas surrounding these galaxies. However, this technique can lead to the selection of interlopers, such as dusty galaxies (e.g., \citealt{2012MNRAS.425L..19H}, \citealt{2023arXiv230315431A}), brown dwarfs (e.g., \citealt{2014MNRAS.439.1038W}) or supernovae (e.g., \citealt{2023ApJ...947L...1Y}). Therefore, spectroscopic follow-up is crucial to confirm the nature of high-redshift galaxy candidates. 

Because of the abundance of ionising photons from primeval galaxies and AGN in the early Universe, Ly$\alpha$ is expected to be very strong at high-redshift (EW$\geq$80\AA, e.g., \citealt{Partridge_1967}, \citealt{1993ApJ...415..580C}) and has been used to spectroscopically confirm some of the most distant galaxies (e.g., \citealt{Zitrin_2015}, \citealt{2016ApJ...823..143R}, \citealt{Jung+22}). 
Several mechanisms are involved in the production of Ly$\alpha$ photons. In the interstellar medium (ISM), young massive stars ionise the neutral hydrogen leading to the emission of Ly$\alpha$ photons by the recombination of hydrogen atoms. This effect is enhanced by the presence of active galactic nuclei (AGN). Further away, in the circumgalactic medium (CGM), Ly$\alpha$ can potentially be excited in shock-heated gas associated with accreting flows. Similarly, outflows can result in  Ly$\alpha$ emission in shock-heated gas. Resonant scattering is also an important mechanism for producing Ly$\alpha$ photons since neutral hydrogen is optically thick to these photons (for a review see \citealt{Ouchi_2020}). In addition to the mechanisms discussed above, the formation of Ly$\alpha$ halos could be reinforced by the presence of nearby companions \citep{Bacon_2021,Leonova_2022}.  Constraining the spatial extent of this strong emission line provides constraints on the ISM and CGM properties,  on the nature of the source (star-forming galaxy or AGN) and on the environment of galaxies, especially those found in the early Universe.  

The large fraction of neutral hydrogen surrounding galaxies in the epoch of reionisation reduces the detectability of Ly$\alpha$ emission originating at the earliest epochs \citep{Robertson_2015, deBarros_2017}. However, within the last decade, several surprising Ly$\alpha$ detections have been obtained from galaxies deep in the epoch of reionisation \citep{Zitrin_2015, Hashimoto+18, bunker_jades_2023}. A recent comparison between JWST observations and simulations proposes an explanation for the detection of Ly$\alpha$ at $z\geq$7 through processes driven by interacting galaxies \citep{Witten+23} located in overdense regions \citep[eg.][]{Leonova_2022, Jung+22, Tang+23, Witstok23}. The most distant Ly$\alpha$ emission line, detected with NIRSpec/JWST data in a HST-selected galaxy \citep{bouwens+2010}, is from GN-z11, at $z=$ 10.6 \citep{bunker_jades_2023} with an integrated flux of 2.3$\times$10$^{-18}$ erg/s/cm$^2$ and an equivalent width (EW) of 18.0$\pm$2.0 $\AA$. Interestingly, the presence of extremely dense gas ($>10^{10}$ cm$^{-3}$, typical of the Broad Line Region of AGN), the detection of [NeIV]$\lambda\lambda$2422,2424 and of CII*1335 emission, indicate that GN-z11 also hosts an AGN  \citep{maiolino_bh_2023}. Spectroscopic follow-up with IFU data has recently provided indications of the presence of PopIII stars in the vicinity of GN-z11 \citep{Maiolino_popIII_23}, making this galaxy one of the most interesting and intriguing objects in the early Universe. 

In this work, we investigate the spatially resolved Ly$\alpha$ emission in the vicinity of GN-z11, the most distant Ly$\alpha$ emitter currently known, by exploiting the recent NIRSpec IFU observations presented in \citet{Maiolino_popIII_23}. 
In section \ref{sec:obs} we briefly describe the observations and the data reduction steps taken to obtain our final reduced datacubes. In section \ref{sec:lya_haloe} we characterise the properties of the Ly$\alpha$ halo and in section \ref{sec:companions} we search for companion galaxies in the local vicinity of GN-z11. Throughout this work, we use the AB magnitude system \citep{1983ApJ...266..713O} and assume the \cite{2020A&A...641A...6P} flat $\Lambda$CDM cosmology. 

%--------------------------------------------------------------------
\section{Observations and data reduction}
\label{sec:obs}

In this work, we use data obtained by the NIRSpec IFU \citep{Jakobsen22,boker22} which is presented in full detail in \citet{Maiolino_popIII_23}. This data was obtained on $22^{\rm nd}-23^{\rm rd}$ of May 2023 under the DDT program 4426 (PI: Roberto Maiolino). The observations used a medium cycling pattern with 10 dithers using a total of 3.3~h on source with the medium-resolution grating/filter pair G140M/F100LP  and 10.6~h with the medium-resolution grating/filter pair G235M/F170LP. This configuration covers 0.97–1.89 $\mu$m and 1.66–3.17 $\mu$m respectively, with a nominal spectral resolution of $R\sim1000$.  The detector was set with the improved reference sampling and subtraction pattern (IRS$^2$), which reduces significantly the readout noise \citep{Rauscher12}.

Despite JWST's excellent pointing accuracy (i.e. 1~$\sigma$ = 0.1$^{\prime\prime}$,  \footnote{\url{https://jwst-docs.stsci.edu/jwst-observatory-characteristics/jwst-pointing-performance}} \citealt{Rigby23}) and accurate target coordinates using GAIA-aligned astrometry, the observations were not centred properly as the guide star selected by the observatory was a binary system, introducing a large offset (1.4$^{\prime\prime}$ towards SE; \citealt{Maiolino_popIII_23}). This offset caused GN-z11 to be close to the edge of the FoV of the NIRSpec/IFU. As a result, for three dither positions, GN-z11 was located at the edge of FoV, while for the remaining seven dithers varying parts of the region west of GN-z11 were outside the FoV of the instrument. This results in steeply decreasing sensitivity of the observations to the West of GN-z11 (and also beyond 0.7$^{\prime\prime}$ North of GN-z11). We re-aligned the astrometry of the NIRSpec/IFU observations to the GAIA-aligned NIRCam. We collapsed the IFU cube in the wavelength range of the NIRCam filter and realigned it with the UV continuum of GN-z11 in the NIRCam observations \citep{tacchella_jades_2023}.

The full data reduction is described in \citet{Maiolino_popIII_23}, here we briefly describe the process. The raw data were processed with the JWST Science Calibration pipeline\footnote{\url{https://jwst-pipeline.readthedocs.io/en/stable/jwst/introduction.html}} version 1.8.2 under CRDS context jwst\_1068.pmap. We made a number of modifications to the reduction steps to improve the data reduction (see  \citealt{Perna23} for more details). The final cubes were created using `drizzle' method setting the spatial pixel scale to  $0.06^{\prime\prime}$ to match the pixel scale of the NIRCam images. For this step, we used an official patch to correct a known bug\footnote{\url{https://github.com/spacetelescope/jwst/pull/7306}}. 

%--------------------------------------------------------------------
\section{Properties of the most distant Ly$\alpha$ halo}
\label{sec:lya_haloe}

\begin{figure*}[ht]%
\centering
\includegraphics[width=0.99\textwidth]{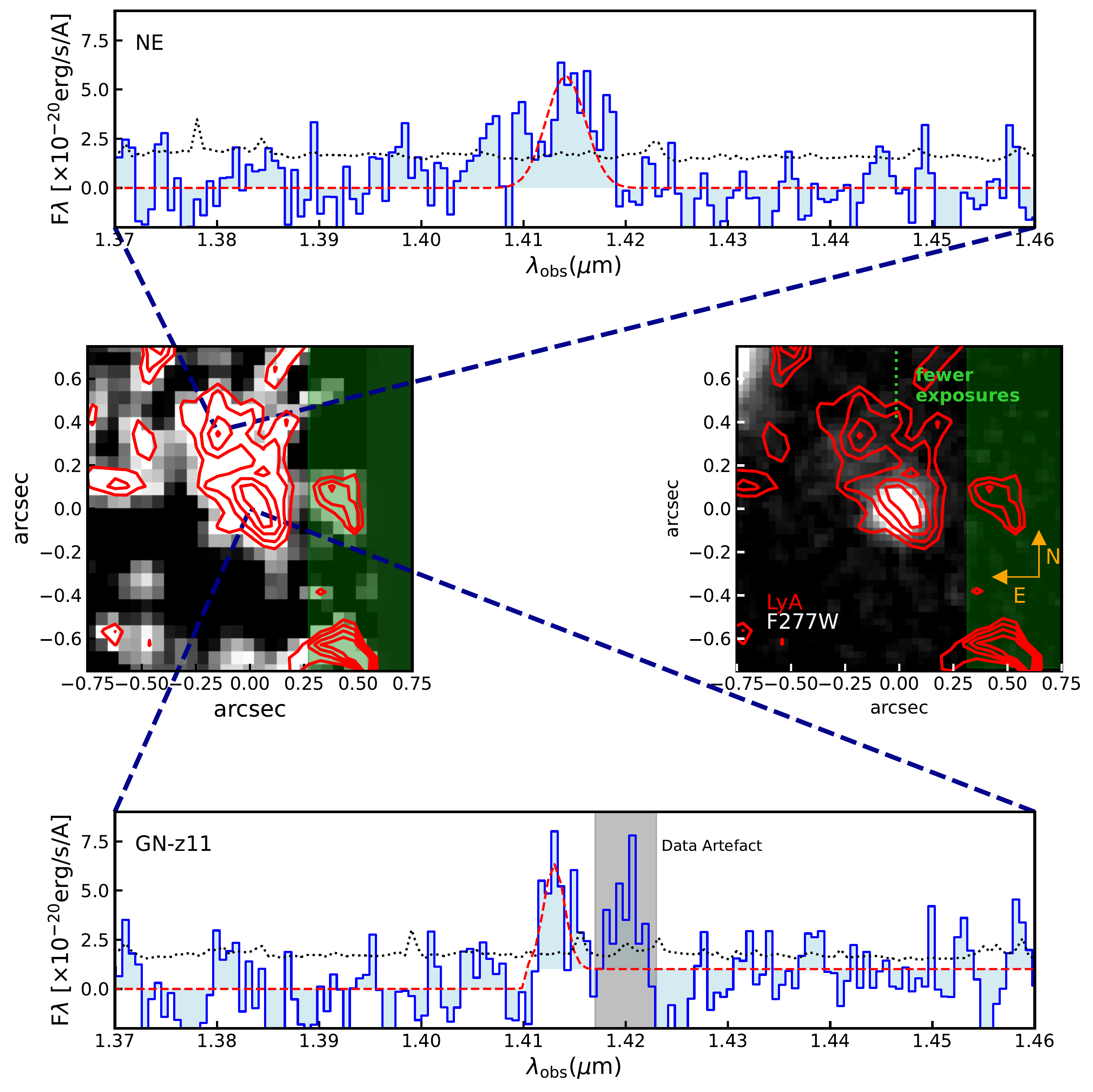}
\caption{Middle row Left: Map of the Ly$\alpha$ emission with red dashed contours (2,3,4 and 5 $\sigma$ levels). Right: JWST/NIRCAM F277W continuum image with Ly$\alpha$ emission map overlaid as red dashed contours (2,3,4 and 5 $\sigma$ levels). The green dotted lines indicate where less than 70\% dither positions cover the area and the green shaded area is covered by  30\% of the dithers. We show extracted spectra from the two Ly$\alpha$ blobs to confirm the Ly$\alpha$ map morphology. The best fit to the emission line is shown as a red dashed line, while the noise is shown as a black dotted line. We highlight the spectral region affected by the data artefact at the edge of the FoV as a grey-shaded region. }\label{fig:Lya_maps}
\end{figure*}

In order to map the individual emission lines we first subtracted the background spectrum. We created the master background by producing a median spectrum of object-free spaxels with a running median filter of 25 channels to avoid any noise spikes. This master background was subtracted from each spaxel in the cube. To create the Ly$\alpha$ map, we collapsed the channels within $\pm 400$ kms$^{-1}$ centred on the peak of the Ly$\alpha$ emission line from \citealt{bunker_jades_2023} (channels 695-703, $\sim 1.4115-1.4152~\mu$m) and subtracted the continuum image (created by taking the channels immediately red-ward of the Ly$\alpha$ emission line and taking the median flux). We smoothed the Ly$\alpha$ image with a Gaussian kernel with a sigma of 1.5 pixels (0.09$^{\prime\prime}$). We show this Ly$\alpha$ image in the left column of Figure~\ref{fig:Lya_maps} as red contours (2, 3, 4 and 5 $\sigma$ levels; with the noise level estimated using the object free central region of the cube) plotted over the NIRCam F277W image. We indicate the region of the data cube with no more than 70\% of dither positions (because of the observatory guiding problem) by a green dotted line and we also highlight the region with $<$30\% of the dither positions by green shaded region. 

The Ly$\alpha$ is extended in the NE direction, with a secondary spatial peak 0.4$^{\prime \prime}$ (i.e.  1.6 kpc) away from the location of the GN-z11 continuum peak, see Figure \ref{fig:Lya_maps}. We also detect an extension of the Ly$\alpha$ to the SW of the continuum centre of GN-z11 on a scale of 0.2 arcseconds, similar to the previous reports of this extension in the MSA observations in \citet{bunker_jades_2023}. The finding of extended CIII]$\lambda$1909 emission in this direction, with a funnel-shaped geometry \citep{maiolino_bh_2023}, indicates that the SW Ly$\alpha$ extension is tracing the AGN ionisation cone. However, the Ly$\alpha$ extension in the latter region should be considered with great care because it is only partially covered by the dithers of the observations and is also potentially subject to field edge issues.

We extracted the spectrum from two separate regions: one encompassing the NE extension (i.e. companion) and the other one centered on the GN-z11 continuum location and also including the SW extension (these regions are shown by blue dashed lines on the NIRCam images in Figure \ref{fig:Lya_maps}). Specifically, we selected these regions to coincide with the structure and peaks of the Ly$\alpha$ map in order to extract the total emission line flux. In the combined 1D spectra, we detect Ly$\alpha$ emission at 4.5$\sigma$ and 6.0$\sigma$ for the central and NE regions, respectively. There is a data artefact around 1.42 $\mu$m, which is only seen in the region where we have 50$\%$ fewer exposures (the SW region of GN-z11), possibly associated with field edge issues \citep[furthermore the second peak hinted by this data artifact is not seen in the much deeper MSA observations,][]{bunker_jades_2023}.

We fitted the extracted spectrum with both a Gaussian profile describing the emission line and a power law with a break at 1.41 microns to describe the galaxy continuum. Although Ly$\alpha$ emission often has a complex line profile, given the low signal-to-noise ratio (SNR) of these observations we fit a simple Gaussian profile. This model was fitted to the data using the Markov chain Monte Carlo (MCMC) algorithm with Python's \texttt{emcee} package \citep{foreman-mackey+2013}. The final quoted parameter values and their uncertainties are the median value and 68\% confidence intervals of the posterior distribution. 

The combined measured flux from the two separate regions (3.9$\times 10^{-18}$ erg s$^{-1}$ cm$^{-2}$) is 70 \% higher than those measured in the MSA observations (\citealt{bunker_jades_2023}). Since flux calibrations of the MSA and IFU observations are accurate to $<5$\% for both instrument modes (\citealt{Boker23}), two factors are likely affecting the MSA measurements: 1) potential self-subtraction of the extended emission line in the MSA observations and 2) the Ly$\alpha$ emission being more extended than the size of the MSA shutter. We summarise Ly$\alpha$ emission line results for each region in Table~\ref{tab:regions}.

The Ly$\alpha$ emission line profile varies significantly across the two separate regions. The FWHM of the Ly$\alpha$ emission line varies between 440 and 1000 km s$^{-1}$ across the regions. Furthermore, the velocity offset of the Ly$\alpha$ emission with respect to the rest-frame optical emission lines from \citet{bunker_jades_2023} ($z=$ 10.6034) is 520 $^{+66}_{-55}$ and 915$^{+86}_{-85}$ km s$^{-1}$ for the central and NE blobs, respectively. We report the results of fitting the Ly$\alpha$ emission from these two regions in Table \ref{tab:regions}. We recover a velocity offset of the Ly$\alpha$ emission line in the local vicinity of GN-z11 that is consistent with that measured by \citet{bunker_jades_2023} and observe an increasing velocity offset in the NE direction, as shown in Figure \ref{fig:Lya_maps}. We see a tentative detection of the unresolved CIV$\lambda\lambda$1548,1551 emission line doublet in the location of the NE blob with a 420 km s$^{-1}$ velocity offset compared to the systemic redshift of GN-z11 (see Appendix for more information). This implies that both GN-z11 and the NE blob have a similar velocity offset of their Ly$\alpha$ emission from their respective systemic redshifts of $\sim 500$ km s$^{-1}$. This large velocity offset is indicative of a very low escape fraction of Lyman-continuum photons ($\sim 0\%$) \citep{Izotov+18, Gazagnes+20, Kakiichi+21}, which is consistent with the escape fraction derived by \cite{bunker_jades_2023} ($3.8\%$).

We also note that the Ly$\alpha$ emission lines of both regions do not show significant asymmetry. While this asymmetry is somewhat typical of high-redshift Ly$\alpha$ emitters \citep{2013Natur.502..524F, 2015ApJ...804L..30O, 2022arXiv221209850J}, \cite{Kakiichi+21} show that with no or few holes in the neutral hydrogen surrounding LyC leaking regions, symmetrical Ly$\alpha$ profiles can be observed. Moreover, \cite{Witten_fesc_LyC_2023} show that the IGM transmission curve can significantly disturb the asymmetry of the Ly$\alpha$ line as it processes through the neutral IGM and hence, the symmetrical Ly$\alpha$ emission line observed here does not necessarily trace the intrinsic profile of the emission line. Furthermore, \cite{bunker_jades_2023} do detect some asymmetry in the Ly$\alpha$ line originating from the position of the UV continuum of GN-z11 and therefore it is possible that the IFU observations do not have the required depth to detect this asymmetry.

\begin{table}
    \centering
    \caption{Measured SNRs, fluxes, FWHM and velocity offsets to the emission lines in GN-z11 for the Ly$\alpha$ emission line extracted from the MSA \citep{bunker_jades_2023} and the regions defined in Figure~\ref{fig:Lya_maps} and \ref{fig:lya_blob}.}
    \begin{tabular}{lcccc}
    % \hline

    \hline
    Region & SNR & Flux & FWHM & $\delta$v \\
           &     & ($\times 10^{-19}$) erg s$^{-1}$ cm$^{-2}$  & km s$^{-1}$&  km s$^{-1}$ \\
    \hline
    MSA$^{*}$ & 12.1 & 23.0$\pm 1.9$ & 530 $\pm 65$ & 555 $^{+32}_{-32}$  \\
    Central & 4.5 & 12.5$^{+3.3}_{-2.8}$ & 440 $^{+204}_{-95}$ & 520 $^{+66}_{-55}$ \\
    NE   & 6.0 & 26.8$^{+4.4}_{-4.5}$ &1000 $^{+145}_{-142}$ & 915$^{+86}_{-85}$\\
    E   & 4.5 & 3.34$^{+0.81}_{-0.75}$ &520 $^{+120}_{-90}$ & 810$^{+66}_{-55}$\\
    %SW  & 3.5 & 16.2$^{+5.2}_{-5.1}$ &580 $^{+250}_{-162}$ &  2167\\
    %SW  & -& <14.0 & - &  -\\
    \hline
    \end{tabular}
    Notes:$^{*}$ Values taken from \citet{bunker_jades_2023} for the full shutter extraction spectrum.
    \label{tab:regions}
\end{table}

The morphology of Ly$\alpha$ emission around GN-z11 is complex with two extended connected peaks. As such it is difficult to assess whether this is a single Ly$\alpha$ halo or two overlapping halos around two separate sources. As a result, we define the Ly$\alpha$ halo size in two different ways: 1) we assume that these are two independent Ly$\alpha$ halos and here we report the size of the halo centred on UV continuum of GN-z11 \citep[similar to the definition from][]{bunker_jades_2023} which would result in a radius of 0.8 kpc; 2) we assume that the entire structure is single asymmetric Ly$\alpha$ halo encompassing the two peaks and we define the size from the UV continuum centre to the furthermost 2$\sigma$ contours of 0.78$\pm$0.1$^{\prime \prime}$ (3.2$\pm$0.4 kpc). 

Before the arrival of the JWST, one of the best instruments to study the profile of Ly$\alpha$ halos at high redshift was the Multi Unit Spectroscopic Explorer (MUSE) at the Very Large Telescope (VLT) \citep{2010SPIE.7735E..08B}. However, its wavelength coverage is not sufficient to detect Ly$\alpha$ halos at $z\geq6.7$. \citet{Leclercq_2017} studied the halos of 145 galaxies at 3 $< z <$ 6 and concluded that 
80\% of their sources have Ly$\alpha$ emission more extended than the UV continuum. \citet{Farina19} presented VLT/MUSE observations of 31 quasars (z=5.7--6.6) with 39\% of the sources revealing Ly$\alpha$ halos. The UV luminosity of GN-z11 (M$_{\rm UV}$ = -21.5) would be considered a bright star-forming galaxy in \citet{Leclercq_2017} and \citet{2018Natur.562..229W} and a faint quasar in \citet{Farina19}, hence, it is necessary to compare the Ly$\alpha$ halo properties to both star-forming galaxies and quasars. 

\begin{figure}[ht]%
\centering
\includegraphics[width=0.45\textwidth]{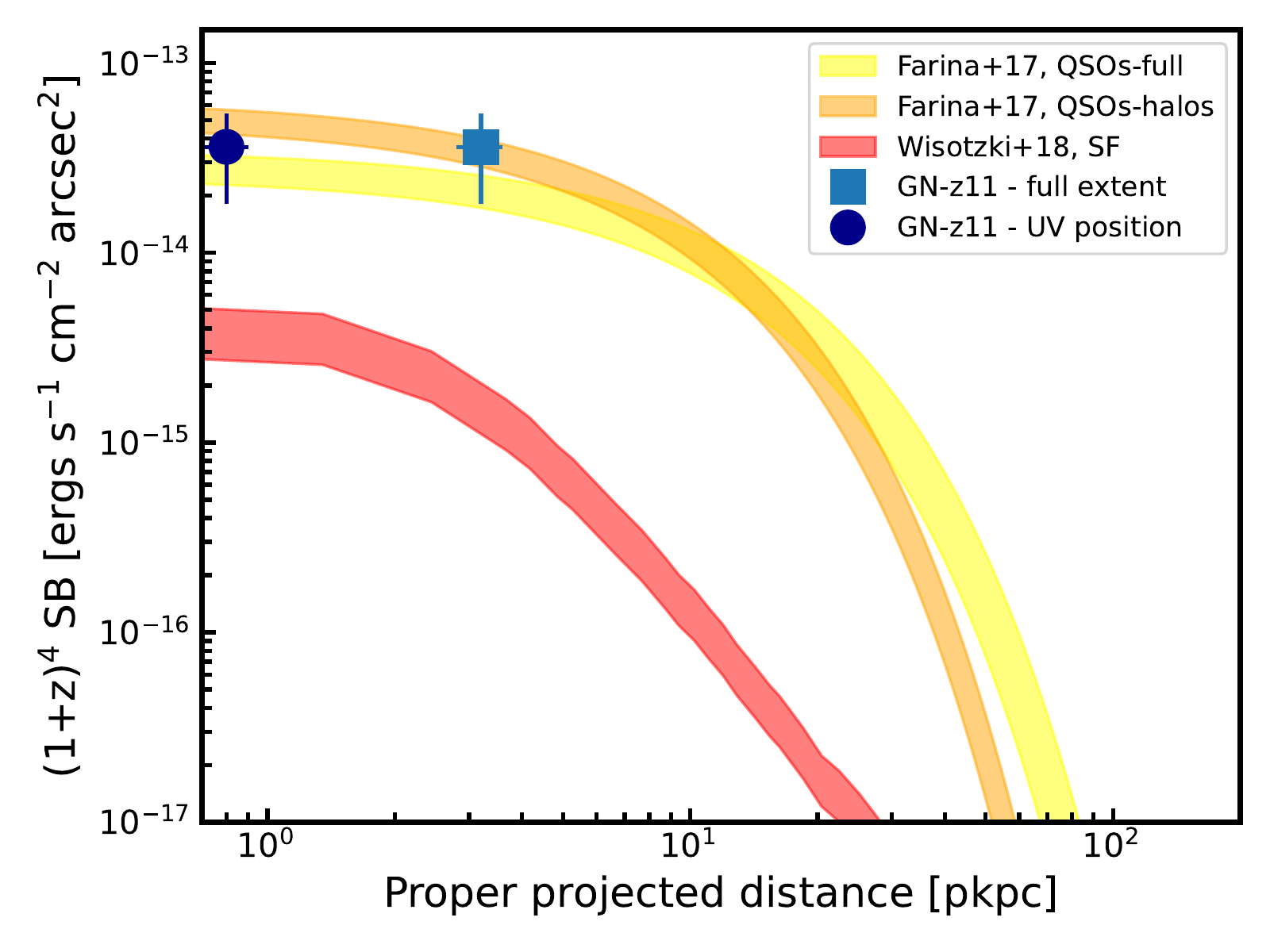}
\caption{ Comparison of the Ly$\alpha$ halo surface brightness of GN-z11 to the profiles of Ly$\alpha$ halos around star-forming galaxies at z=5.0-6.0 \citep[red shaded region][]{2018Natur.562..229W} and quasars z=5.7-6.6 \citep[yellow and orange shaded region][]{Farina19}. For the quasars from \citet{Farina19} we show the radial surface brightness profiles for the full sample and for only those with detected Ly$\alpha$ halos as yellow and orange regions, respectively. We show two separate definitions of halo size for GN-z11: a smaller halo in the vicinity of UV continuum (dark blue circle) and the full extent of the asymmetric halo (light blue square). The surface brightness and the extent of GN-z11's Ly$\alpha$ halo are in agreement with quasars rather than star-forming galaxies at z$\sim$6 . } 
\label{fig:profile_radius}
\end{figure}

We compare the surface brightness of GN-z11 to the profiles of star-forming galaxies from \citet{2018Natur.562..229W} and quasars from \citet{Farina19} in Figure \ref{fig:profile_radius}. To account for cosmological dimming across different redshifts, we multiply the surface brightness profiles by (1+z)$^4$ factor to make an easier comparison. We show two separate measurements for GN-z11 based on the definitions outlined above; 1) a single smaller halo centre on GN-z11 UV continuum (dark blue circle); 2) the extent of the full larger halo encompassing both peaks (light blue square). Remarkably, the size and surface brightness of the GN-z11 halo is consistent with those around quasars rather than star-forming galaxies. We also note that the asymmetric  Ly$\alpha$ halo currently observed around GN-z11, although potentially as a result of insufficient dithers due to the guiding problem, does mimic the structure of Ly$\alpha$ halos around quasars in overdense regions at lower redshifts \citep[e.g.][]{Cantalupo+14, Hennawi+15, Bacon+21}. 

We also compare our measurements with what is expected from zoom-in simulations of a large halo of $M_\mathrm{h}=6.9\times$10$^{12}$\,M$_{\odot}$ at $z=6$ \citep{Bennett+23, Sijacki_2009}, with variations in the feedback model from the \textsc{fable} suite of simulations \citep{Henden_2018}. These simulations are a higher resolution version of those described in \citet{Bennett+23}, and use the same physical models. \citet{bunker_jades_2023} estimate a metallicity for GN-z11 of 0.12\,Z$_{\odot}$, which corresponds to a dust-to-metal ratio (DMR) of $\sim$0.01 \citep{Li_2019}. Our Ly$\alpha$ modelling uses the \textsc{colt} code \citep{ASmith_2015}, which includes Ly$\alpha$ emission from unresolved HII regions around stars and the effect of ionisation from the AGN on the surrounding gas, and we use two different DMRs of 0.01 and 0.1. Our simulations conclude that (i) the size of the halo depends on the DMR (the halo is 25-33\% larger with a DMR=0.01 than with a DMR=0.1) and (ii) the expected size of the GN-z11 halo ranges between 2.5 to 3.0 kpc. 

%--------------------------------------------------------------------
%--------------------------------------------------------------------

%--------------------------------------------------------------------
\section{Search for companions in the vicinity of GN-z11}
\label{sec:companions}

The presence of nine galaxies within an area of $\sim 100$ Mpc$^2$ with photometric redshifts within $\Delta z$=1 of the redshift of GN-z11, places GN-z11 at the centre of a clustered region, reported by \cite{tacchella_jades_2023}. This could provide an explanation for the escape of Ly$\alpha$ radiation through the intervening neutral IGM, via the production of a large ionised bubble \citep{Tilvi_2020, Leonova+22, Witstok23}. Moreover, the presence of a HeII clump, reported by \cite{Maiolino_popIII_23}, that may host PopIII stars, provides the first evidence of a potential companion galaxy just $0.6^{\prime\prime}$ from GN-z11. The presence of such companions helps to explain the production and escape of Ly$\alpha$ radiation through the local environment of high-redshift galaxies \citep{Witten+23}. 

With this in mind, we search for emission lines in the NIRSpec IFU FoV that can be attributed to companion galaxies at a redshift consistent with that of GN-z11. 

We note that no photometric candidate companions exist within the limited FoV of these observations. Any such companions identified with these observations will therefore have no associated UV continuum. However, \cite{Kerutt+22} found that one-third of LAEs in MUSE have no associated UV continuum counterpart and hence this would not be unexpected. 

\begin{figure*}%
\centering
\includegraphics[width=0.9\textwidth]{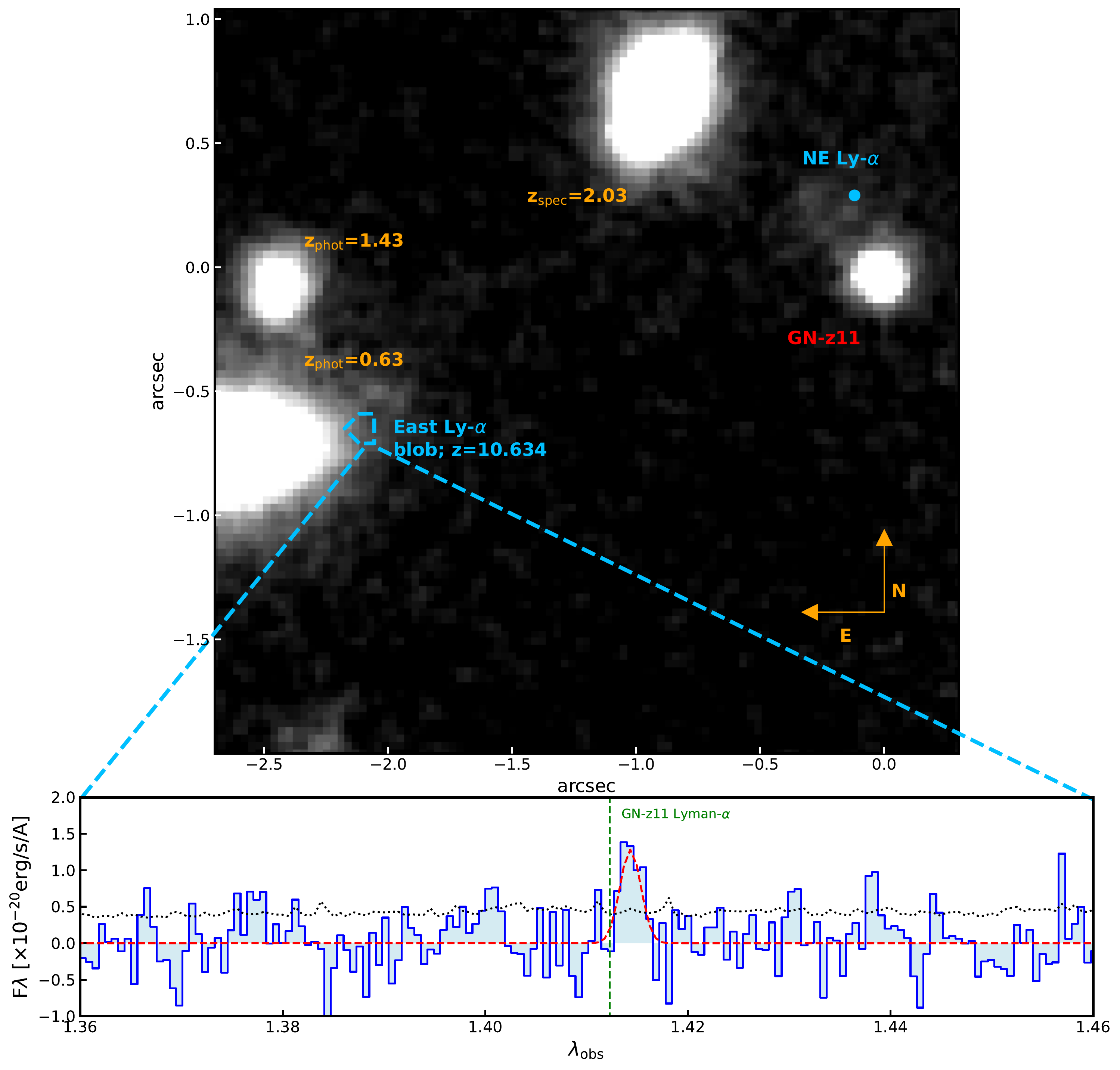}
\caption{Detection of an additional Ly$\alpha$ emitter, 2.1 arcseconds to the East of GN-z11. Top: JWST/NIRCam F277W image corresponding to the FoV of our NIRSpec/IFU observations. We indicate the region from which we extracted the spectrum on the right with blue dashed contours. We also note three foreground galaxies visible in the NIRCam image with their corresponding redshifts. Bottom: Extracted Ly$\alpha$ spectrum from the region indicated in the NIRCam image. The red dashed line shows the best fit to the spectrum while the black dotted line shows the error spectrum. The green dashed line shows the wavelength of Ly$\alpha$ emission seen in the main component of GN-z11. The foreground galaxies in the vicinity show no visible emission lines that would contaminate this spectrum.} \label{fig:lya_blob}
\end{figure*}

\subsection{\texorpdfstring{Ly$\alpha$}{Lya} emitting blobs}

Figure~\ref{fig:Lya_maps} reveals a Ly$\alpha$ ``blob” $\sim 0.4^{\prime\prime}$ to the north-east of GN-z11, originating just above the ``haze” reported by \cite{tacchella_jades_2023}. This 6 $\sigma$ detection of Ly$\alpha$ spectroscopically confirms the presence of a companion galaxy.
Indeed, we also have a tentative CIV$\lambda$1550 detection in the NE Ly$\alpha$ blob at 3.2$\sigma$ (see Appendix \ref{sec:CIV}) 
indicating a potentially highly ionising source in that region, such as a very young stellar population or a faint AGN \citep[see][for more details]{Maiolino_popIII_23}.

Moreover, the detection of HeII emission from a clump $\sim 0.6^{\prime\prime}$ to the north-east of GN-z11, originating from a distinct region to the Ly$\alpha$ halo \citep{Maiolino_popIII_23} indicates the presence of pristine gas, possibly photoionised by a cluster of PopIII stars. 

This detection of emission lines originating from clumps that are distinct from GN-z11 offers a deeper insight into the processes driving Ly$\alpha$ emission at these epochs. It acts as further support for the idea that galaxy interactions are driving Ly$\alpha$ emission in the epoch of reionisation \citep{Witten+23}, and confirms that a significant amount of neutral hydrogen around these sources must have been displaced in order to facilitate the observation of Ly$\alpha$ photons from both sources, despite their different properties. This result also suggests that peculiar velocities may not be playing a significant role in the escape of Ly$\alpha$ photons given that both GN-z11 and the NE region have relatively large velocity offsets from each other ($\sim 400$ km/s). If a large velocity offset driven by gravitational interactions between GN-z11 and these clumps was driving the escape of Ly$\alpha$ photons through the neutral IGM, it could be expected that only one source would show Ly$\alpha$ emission. That source would emit Ly$\alpha$ photons that are heavily redshifted relative to the neutral IGM and hence, are more likely to escape without attenuation, while the other source would emit Ly$\alpha$ photons that are blueshifted and hence would be absorbed by the neutral hydrogen in the IGM unless a significant ionised bubble is present \citep[eg.][]{Tilvi_2020, Leonova+22, Witstok23, Witten+23}. While the interaction between GN-z11 and its companions is more complex than a simple two-body interaction, this result offers an early indication that peculiar velocities may not be the most important factor in facilitating Ly$\alpha$ escape. 

We also searched for more distant Ly$\alpha$-emitting sources in the full FoV of the NIRSpec IFU observations (specifically the area fully covered by the 10 dithers) within $\pm2000$ km s$^{-1}$ of the systematic redshift of GN-z11. We find a single candidate detected 2.15 arcseconds East of GN-z11 and we show the location of the candidate and its spectrum in Figure \ref{fig:lya_blob}. The emission  line is detected at 4.5$\sigma$ with FWHM of 520$^{+120}_{-90}$ km s$^{-1}$ at z=10.6346, 290 km s$^{-1}$ offset redwards of the Ly$\alpha$ emission line of GN-z11. 

Since the Ly$\alpha$ emitter is close to the foreground galaxy, we verified that the emission line is indeed from a $z\sim$10.63 galaxy and not the foreground source \citep[$z_{\rm phot}$=0.63;][]{Hainline23}. We extracted a spectrum with a radius of 0.3$^{\prime\prime}$ centred on the source East of the Ly$\alpha$ blob in both band 1 and band 2 IFU observations. Although the continuum is well detected in both bands, we do not detect any emission lines in this foreground source. Assuming that the Ly$\alpha$ emission from the candidate is in fact an emission line coming from the foreground source at $z\sim0.63$, the only possible emission line would be CaII$\lambda$8664, which we do not expect to see without detection of the [SIII]$\lambda$9532, HeI$\lambda$10830 and Pa$\delta$ emission lines. 

Given that these emission lines are undetected, we conclude that this emission is not coming from the foreground galaxy, but is indeed an additional Ly$\alpha$ emitter at the redshift similar to GN-z11.

\subsection{A candidate $z=10.6$ protocluster core}

\cite{tacchella_jades_2023} previously identified a photometric candidate overdensity surrounding GN-z11 within {\bf a $\sim$ 10 cMpc $\times$ 10 cMpc region}, which corresponds to $\sim$430 pkpc at z=10.6. The detection of the HeII blob \citep{Maiolino_popIII_23}, the Ly$\alpha$ blob $\sim 0.4^{\prime\prime}$ NE of GN-z11 and the detection of a further Ly$\alpha$ blob 2.1$^{\prime\prime}$ to the East brings the number of spectroscopically confirmed $z\sim10.6$ objects in the NIRSpec IFU FoV to four. {\bf Previous theoretical works \citep[eg.][]{Chiang+17} have demonstrated that the expected size of a protocluster core is already on scales of just hundreds of ckpc by $z\sim7$. Observations of protocluster cores at a similar redshift \citep[eg.][]{Capak+11, Arribas23, Hashimoto+23, Morishita+23} have also found a central core of order 150 ckpc. This finding motivates our conclusion that the objects seen within the IFU FoV (140 ckpc $\times $ 140 ckpc) are representative of the core of a protocluster. Moreover, theoretical work by \cite{Chiang+17} has shown that the overall size of the protocluster at high redshifts like these is typically of order 10 cMpc, consistent with the larger scale overdensity of galaxies identified by \cite{tacchella_jades_2023}}.

Comparing the number of objects identified in the IFU FoV at $z=10.6$ we find an overdensity parameter of $\delta \geq$27. We estimate the number of galaxies that are expected in the FoV by integrating the luminosity function of \cite{Harikane23}, at $z\sim$10, down to a UV magnitude that is determined by assuming that the Ly$\alpha$ blobs have a maximum Ly$\alpha$ rest-frame equivalent width of $\sim 1000 \rm{\AA}$. We make this assumption as these Ly$\alpha$ blobs have no associated UV continuum. We give the overdensity parameter as a lower bound as the FoV of the NIRSpec IFU is too small to accurately constrain the overdensity parameter, and thus we use the upper-bound expected number of galaxies to give us an absolute lower bound on our overdensity parameter. This value is significantly larger than the overdensity parameter for the $z = 7.66$ protocluster core presented in \cite{Laporte+22} and, following the same method as above, we estimate an overdensity parameter of $\delta \geq$11 for the $z=7.88$ protocluster core presented in \cite{Hashimoto+23}. Given that the region surrounding GN-z11 is more overdense than previously observed protocluster cores at $z\sim 8$, that it resides in a large-scale photometric redshift overdensity \citep{tacchella_jades_2023} and that protoclusters are expected to fuel AGN in their central galaxy \citep[eventually forming the observed population of massive $z\sim6$ quasars, see][]{Bennett+23}, as is observed in the case of GN-z11 \citep{maiolino_bh_2023}, the region surrounding GN-z11 appears to be a strong candidate for a protocluster core.Moreover, GNz11 is by far the brightest galaxy in this structure, potentially making it a precursor of the brightest cluster galaxies seen at lower redshifts.

\begin{figure}[h]%
\centering
\includegraphics[width=0.5\textwidth]{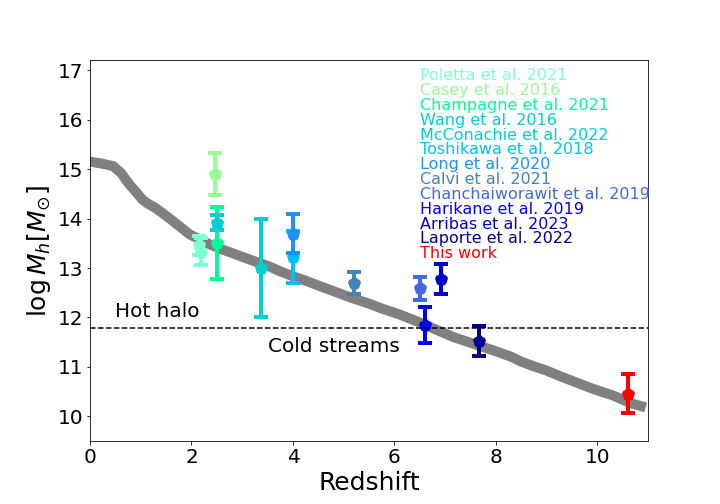}
\caption{Evolution of the protocluster dark matter halo mass as a function of redshift. The grey line shows the evolution of a protocluster leading to a Coma-like cluster at $z=0$ \citep{2013ApJ...779..127C}. We extrapolate this evolution to $z\sim$11 assuming no change in the slope from $z\sim4$. The dark matter halo mass of GN-z11 is fully consistent with this evolution. } 
\label{fig.evolution}
\end{figure}

As a preliminary analysis, we estimate the dark matter halo mass of this structure following \citet{Behroozi2013}. We assume that the total stellar mass of this protocluster core is dominated by GN-z11 and obtain a dark matter halo mass of $M_\mathrm{h}$=2.96$^{+0.44}_{-0.39} \times$10$^{10}$\,M$_{\odot}$. This is a reasonable assumption given that all objects within the IFU field-of-view do not have any UV counterpart in the NIRCam images. Moreover, candidates identified in \citet{tacchella_jades_2023} are 3 magnitudes fainter than GN-z11 and will therefore have much smaller stellar masses than GN-z11. We compare this dark matter halo mass with previous findings at lower redshift (\citealt{2021A&A...654A.121P},  \citealt{2016ApJ...824...36C}, \citealt{2021ApJ...913..110C}, \citealt{2016ApJ...828...56W}, \citealt{2022ApJ...926...37M}, \citealt{2018PASJ...70S..12T}, \citealt{2020ApJ...898..133L}, \citealt{2021MNRAS.502.4558C}, \citealt{2019ApJ...877...51C}, \citealt{2019ApJ...883..142H}, \citealt{Laporte+22}) and with the expected evolution of a Coma-like cluster at $z=0$ \citep{2013ApJ...779..127C} assuming a constant slope in the evolution at  $z\geq$4 (see Figure \ref{fig.evolution}). The estimated dark matter halo mass of the GN-z11 protocluster is fully consistent with what is expected for a Coma-like cluster progenitor at $z=$10.6. 

%--------------------------------------------------------------------
%--------------------------------------------------------------------

\section{Conclusions}

We have presented an analysis of JWST/NIRSpec IFU observations of GN-z11, an extremely luminous galaxy hosting an AGN at $z=$10.603. Due to a telescope guiding problem, GN-z11 is near the edge of the FoV and hence we are only able to map the Ly$\alpha$ halo extension on and to the North and East of GN-z11.

We detect extended Ly$\alpha$ emission at the location of GN-z11 and towards the NE. Extracting the spectrum from the two separate regions (GN-z11 and the NE extension) we have measured a velocity offset of the Ly$\alpha$ emission with respect to GN-z11 of 520 $^{+66}_{-55}$ and 915$^{+86}_{-85}$ km s$^{-1}$ for the central and NE extension, respectively. We have measured the extent of the Ly$\alpha$ halo to be $0.8-3.2$kpc; based on two separate definitions of the size (i.e. edge to edge of the whole structure or the size of the smaller-scale halo around the UV continuum of GN-z11). Comparing the radial surface brightness of GN-z11 to previous studies at z=5-6.5 \citep{2018Natur.562..229W, Farina19} shows that the Ly$\alpha$ halo is consistent with those of quasars rather than star-forming galaxies. 

We also detect a distinct Ly$\alpha$ blob to the NE of GN-z11. Together with the previously identified HeII clump \citep{Maiolino_popIII_23}, this indicates the presence of multiple companions surrounding GN-z11. This appears to support the theory that companion galaxies are playing a key role in the detectability of Ly$\alpha$ emitting galaxies deep into the epoch of reionisation \citep{Witten+23}. 

The further identification of a Ly$\alpha$ blob 2.1$^{\prime\prime}$ East of GN-z11 brings the number of objects within the NIRSpec IFU FoV (3$^{\prime\prime}$x3$^{\prime\prime}$) to four, making this region more overdense than previously discovered protocluster cores at $z\sim8$. This significant overdensity, combined with the larger scale overdensity reported in \citet{tacchella_jades_2023} and the presence of an AGN in GN-z11 \citep{maiolino_bh_2023}, at the centre of this system, makes the region studied in this work a strong candidate for a protocluster core. Our preliminary analysis of this protocluster demonstrates that it has a dark matter halo mass consistent with what is expected for a Coma-like cluster progenitor at $z \sim 10.6$. Our study confirms that the most massive galaxies identified with HST, like GN-z11, are indeed surrounded by much fainter, previously undetected, galaxies at similar redshifts which likely form the first large-scale structure in the Universe.   

\begin{acknowledgements}
The authors would like to thank Yucheng Guo for a very productive and helpful discussion.
FDE, JS, RM, TL, WB acknowledge support by the Science and Technology Facilities Council (STFC), ERC Advanced Grant 695671 ``QUENCH". RM also acknowledges funding from a research professorship from the Royal Society.
CW thanks the Science and Technology Facilities Council (STFC) for a PhD studentship, funded by UKRI grant 2602262.
NL acknowledges support from the Kavli foundation. 
H{\"U} gratefully acknowledges support by the Isaac Newton Trust and by the Kavli Foundation through a Newton-Kavli Junior Fellowship.
JSB acknowledges support from the by the Simons Foundation Collaboration on “Learning the Universe”.
M.P., S.A and B.R.P. acknowledge support from the research project PID2021-127718NB-I00 of the Spanish Ministry of Science and 
Innovation/State Agency of Research (MICIN/AEI/ 10.13039/501100011033), and M.P. the Programa Atraccion de Talento de la Comunidad de 
Madrid via grant 2018-T2/TIC-11715
AJB, acknowledge funding from the "FirstGalaxies" Advanced Grant from the European Research Council (ERC) under the European Union’s Horizon 2020 research and innovation programme (Grant Agreement No. 789056)
S.C and G.V. acknowledge support by the European Union’s HE ERC Starting Grant No. 101040227 - WINGS.
ECL acknowledges the support of an STFC Webb Fellowship (ST/W001438/1)
DJE is supported as a Simons Investigator and by JWST/NIRCam contract to the University of Arizona, NAS5-02015
BER acknowledges support from the NIRCam Science Team contract to the University of Arizona, NAS5-02015. 
RS acknowledges support from an STFC Ernest Rutherford Fellowship (ST/S004831/1).
The research of CCW is supported by NOIRLab, which is managed by the Association of Universities for Research in Astronomy (AURA) under a cooperative agreement with the National Science Foundation.
C N A.W. acknowledges JWST/NIRCam contract to the University of Arizona NAS5-02015
The authors acknowledge the use of the lux supercomputer at UC Santa Cruz, funded by NSF MRI grant AST 1828315.
The simulations described in this work were performed at the DiRAC@Durham facility managed by the Institute for Computational Cosmology on behalf of the STFC DiRAC HPC Facility (www.dirac.ac.uk). The equipment was funded by BEIS capital funding via STFC capital grants ST/P002293/1, ST/R002371/1 and ST/S002502/1, Durham University and STFC operations grant ST/R000832/1. DiRAC is part of the National e-Infrastructure.
\end{acknowledgements}

% WARNING
%-------------------------------------------------------------------
% Please note that we have included the references to the file aa.dem in
% order to compile it, but we ask you to:
%
% - use BibTeX with the regular commands:
%   \bibliographystyle{aa} % style aa.bst
%   \bibliography{Yourfile} % your references Yourfile.bib
%
% - join the .bib files when you upload your source files
%-------------------------------------------------------------------

\bibliographystyle{aa}
\bibliography{mybib}

\section{Appendix}
\subsection{Tentative detection of CIV in NE blob}\label{sec:CIV}

\begin{figure}[h]%
\centering
\includegraphics[width=0.45\textwidth]{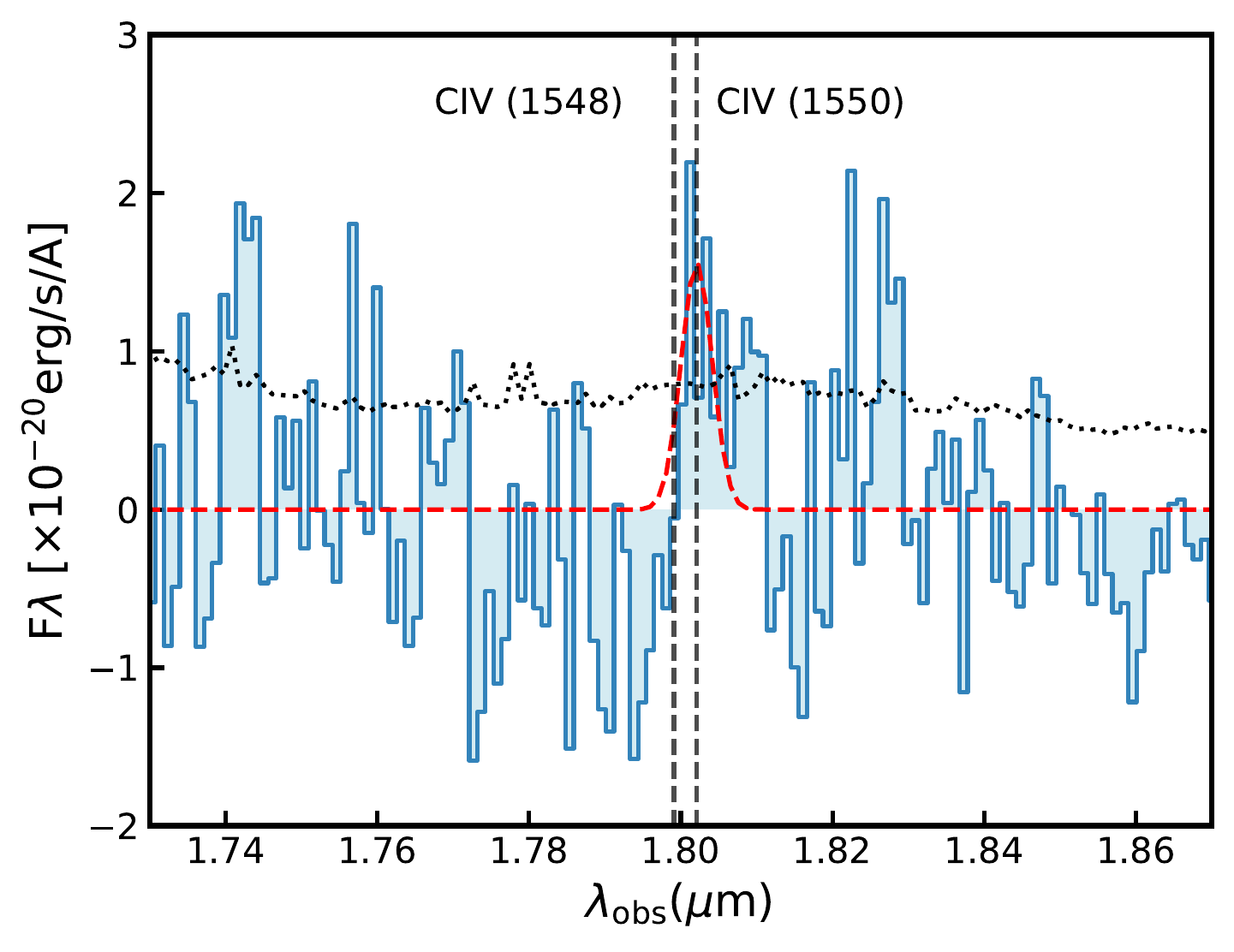}
\caption{Tentative detection of CIV$\lambda\lambda$1548,1550 in the location of NW Ly$\alpha$ blob. The red dashed line indicates the best fit to the CIV$\lambda$1551 line and the black dotted lines show the noise spectrum.} \label{fig:CIV}
\end{figure}

As we discussed in the \S \ref{sec:companions} we search for any other emission line at the same locations as the Ly$\alpha$ blobs detected with NIRSpec/IFU observations. We see a tentative detection of CIV$\lambda$1551 emission within the same aperture as the bottom panel of Figure \ref{fig:Lya_maps}. We show the extracted spectrum and the best fit in Figure \ref{fig:CIV}. The line is detected at 3.2$\sigma$ in the Band 2 observations at z= 10.620 (i.e. +420 km s$^{-1}$ of GN-z11) with a flux of $7.54^{+3.12}_{-2.35} \times 10^{-19}$ ergs s$^{-1}$ cm$^{-2}$ and a FWHM of $730^{+290}_{-240}$ km s$^{-1}$.

The detection of CIV requires very hard ionisation field mostly likely coming from young high-mass stars. However, this tentative detection needs to be confirmed by the future re-observations of this programme.

\end{document}